\documentclass[twocolumn,showpacs,pre,aps]{revtex4}

\usepackage{dcolumn}
\usepackage{bm}

\begin{document}

\title{Designing Langevin Microdynamics in Macrocosm}

\author{Yuriy E. Kuzovlev}
\email{kuzovlev@kinetic.ac.donetsk.ua}
\affiliation{A.A.Galkin Physics and Technology Institute
of NASU, 83114 Donetsk, Ukraine}

\date{\today}

\begin{abstract}
Previously developed ``stochastic representation of deterministic
interactions`` enables exact treatment of an open system without
leaving its native phase space (Hilbert space) due to peculiar
stochastic extension of Liouville (von Neumann) equation for its
statistical operator. Can one reformulate the theory in terms of
stochastic ``Langevin equations'' for its variables? Here it is shown
that in case of classical Hamiltonian underlying dynamics the answer
is principally positive, and general explicit method of constructing
such equations is described.
\end{abstract}


\maketitle

{\,\bf I. Introduction.\,} Any Langevin equations involve
irreversibility (friction) and indeterminism (noise), as the
classical equations which imitate interaction between ``Brownian
particle'' and a fluid (see e.g. \cite{isi} and references therein).
Both the friction and noise represent the same reversible and
deterministic microscopic dynamics, but usually are presumed
unambiguously (additively) distinguishable. In general, of course,
such assumption is wrong, because the friction itself can essentially
fluctuate, as in the case of interaction between macroscopic
vibrations of a quartz crystal and its own phonon gas (see e.g.
\cite{i0} and references therein). Therefore the question arises: how
one should construct "Langevin equations" (interpreted loosely as a
model replacement of underlying microscopic dynamics) to be sure they
result quite accurate and thus free of artifacts?

The answer can be formulated in the framework of ``the stochastic
representation of deterministic interactions''
\cite{i1,i2,i3,i4,i5,i6}, at least in two widespread situations:

i)\, when the dynamics is Hamiltonian while interaction between a
system of interest, ``D'', and other world, ``B'', is described by a
bilinear contribution to Hamiltonian of ``D+B'' \cite{i1,i2,i3,i5} :
\begin{equation}
\begin{array}{c}
H=H_d+H_b+H_{int}\,\,,\,\,\, H_{int}=\sum_n D_n B_n\,\,; \label{blh}
\end{array}
\end{equation}
the marks ``d'' and ''b'' and the operators (or phase functions, in
classical mechanics) $\,D_n\,$ and $\,B_n\,$ relate to ``D'' and
``B'', respectively;

ii)\, when joint evolution operator of ``D+B'', $\,L\,$, has similar
bilinear form \cite{i3,i4} :
\begin{equation}
\begin{array}{c}
L=L_d+L_b+L_{int}\,\,,\,\,\, L_{int}=\sum_n \Lambda ^{d}_{n}\Lambda
^{b}_{n}\,\, \label{bll}
\end{array}
\end{equation}

The evolution operator is understood as those governing join
statistical operator, $\,\rho\,$ , of ``D+B'':
\begin{equation}
\begin{array}{c}
d\rho/dt=L\rho\,\,\label{ev}
\end{array}
\end{equation}
In Hamiltonian dynamics, $\,L=\mathcal{L}(H)\,$, where
$\,\mathcal{L}(H)\,$ is quantum or classical Liouville operator,
\[
\begin{array}{c}
\mathcal{L}(H)\rho=\frac
{i}{\hbar}[\rho\,,H]\,\,\,\,\,\text{or}\,\,\,\,\,
\mathcal{L}(H)\rho=\left(\frac {\partial H}{\partial q}\frac
{\partial}{\partial p}-\frac {\partial H}{\partial p}\frac
{\partial}{\partial q}\right)\rho\,\,,
\end{array}
\]
with $\,\{q,p\}\,$ being canonic variables. In case (\ref{blh})
$\,L=\mathcal{L}(H)\,$ always has the bilinear form (\ref{bll})
\cite{i3,i4} (the case (\ref{bll}) covers also non-canonic treatments
of Hamiltonian dynamics \cite{i4} and, besides, essentially
non-Hamiltonian and  irreversible dynamics, and even Markovian
probabilistic evolutions).

For simplicity, in this paper discussion of the Langevin equations
will be confined by classical mechanics, moreover, starting from
Sec.III, by the case (\ref{blh}) only.

{\,\bf II. Characteristic functionals.\,} The statistical operator
$\,\rho\,$ from Eq.\ref{ev} (density matrix, probability measure,
etc.) says about current state $\,\Gamma =\Gamma_d\oplus\Gamma_b\,$
of ``D+B'' only. Who is interested also in its correlations with its
prehistory, may consider one or another characteristic functional
(CF)
\[
\begin{array}{c}
\text{Tr}_d\,\text{Tr}_b\,\,\rho(t,\Gamma)\left\langle
\exp\,[\,\int_{t>t^{\prime}}\sum_j v_j(t^{\prime})Q_j(\Gamma
(t^{\prime}))\,dt^{\prime}\,]\right\rangle^{\Gamma }\equiv
\end{array}
\]
\begin{equation}
\begin{array}{c}
\equiv\,\left\langle\exp\,[\,\int_{t>t^{\prime}} \sum_j
v_j(t^{\prime})Q_j(t^{\prime})\,dt^{\prime}\,]
\right\rangle\,\,\,,\label{dcf}
\end{array}
\end{equation}
where $\,Q_j(\Gamma )\,$ are some phase functions (i.e. functions of
instant system's state) and $\,v_j(t)\,$ conjugated arbitrary test
functions (probe functions); \,\,Tr$_b\,$ and \,Tr$_d\,$ denote
``traces'' over phase spaces of ``B'' and ``D'', that is integrations
over $\,\Gamma_b\,$ or $\,\Gamma_d\,$; \,$\,\left\langle
...\right\rangle^{\Gamma }\,$ is conditional statistical average
under given present state $\,\Gamma=\Gamma(t)\,$,\, and the
right-hand side retells the left from viewpoint of exterior
observers. Particularly, in case of deterministic dynamics the
conditional averaging degenerates into replacing $\,\Gamma
(t^{\prime})\,$ by strictly definite function of
$\,\Gamma=\Gamma(t)\,$.

In any case, if readdressing symbol $\,\rho \,$ to the whole
expression under the traces in (\ref{dcf}), one can write
\begin{equation}
\begin{array}{c}
\left\langle \exp\,[\,\int_{t>t^{\prime}} \sum_j
v_j(t^{\prime})Q_j(t^{\prime})\,dt^{\prime}\,] \right\rangle
=\,\text{Tr}_d\,\text{Tr}_b\,\,\rho\,\,\,,\label{gcf}
\end{array}
\end{equation}
where now, obviously, $\,\rho \,$ obeys the equation
\begin{equation}
\begin{array}{c}
d\rho/dt=\{\sum_j v_j(t)Q_j(\Gamma) +L\}\,\rho\,\,\,\label{gev}
\end{array}
\end{equation}
instead of (\ref{ev}). Thus one reduces CF to slightly modified
evolution equation. In fact that is a sort of famous relations
between path integrals and differential equations, like the
Feynman-Kac formulas \cite{rs,f}. Nevertheless, we once more accented
the transition from (\ref{dcf}) to (\ref{gcf})-(\ref{gev}) (see also
Sec.2 in \cite{i6}) because, curiously, some referees are not
familiar with such possibility (by the way, some similar old examples
can be found in \cite{bk3,bk2}).

{\,\bf III. Stochastic representation.\,} Consider partial
probability measure of ``D'''s states, $\, \rho_d \equiv
\,$Tr$\,_b\,\rho\,$, where $\,\rho \,$ satisfies the evolution
equation (\ref{ev}). According to \cite{i1,i2,i3}, if once $\,\rho\,$
was factored, then later $\,\rho_d\,$ can be represented as the
average
\begin{equation}
\begin{array}{c}
\rho_d=\left\langle\left\langle
\widetilde{\rho}_d\,\right\rangle\right\rangle\,\, \label{sr}
\end{array}
\end{equation}
of stochastic probability measure $\,\widetilde{\rho}_d\,$ which
obeys the time-local differential equation
\begin{equation}
\frac {d\widetilde{\rho}_d}{dt}=\left[\sum_n y_n(t)D_n
+\mathcal{L}\left(H_d+\sum_n
x_n(t)D_n\right)\right]\widetilde{\rho}_d\,\label{se}
\end{equation}
with $\,x_n(t)\,$ and $\,y_n(t)\,$ being definite stochastic
processes and
$\,\left\langle\left\langle...\right\rangle\right\rangle\equiv
\langle\left\langle...\right\rangle_y\rangle_x\equiv
\left\langle\left\langle...\right\rangle_x\right\rangle_y\,$
statistical average with respect to them. Similarly, if the phase
functions $\,Q_j\,$ wholly belong to ``D'' then their CF (\ref{dcf})
can be represented, in place of (\ref{gcf}) and (\ref{gev}), as
\begin{equation}
\begin{array}{c}
\left\langle \exp\,[\,\int_{t>t^{\prime}} \sum_j
v_j(t^{\prime})Q_j(t^{\prime})\,dt^{\prime}\,] \right\rangle
=\left\langle\left\langle\,\text{Tr}_d\,\,\widetilde{\rho}_d\,
\,\right\rangle\right\rangle\,\,\,,\label{qcf}
\end{array}
\end{equation}
where now $\,\widetilde{\rho}_d\,$ is a solution of the stochastic
equation
\begin{equation}
\frac {d\widetilde{\rho}_d}{dt}=[\,v_j(t)Q_j+ y_n(t)D_n
+\mathcal{L}\left(H_d+
x_n(t)D_n\right)]\,\widetilde{\rho}_d\,\label{se1}
\end{equation}
with the same random sources $\,x_n(t)\,$ and $\,y_n(t)\,$ (closely
repeated indices imply summation).

Notice that (\ref{qcf}) and (\ref{se1}) again exploit the Feynman-Kac
type relations, now for stochastic evolution operator $\,[\,y_n(t)D_n
+\mathcal{L}(H_d+ x_n(t)D_n)\,]\,$ in place of $\,L\,$, and that next
such instants will not commented.

It is easy to see that $\,x_n(t)\,$ surrogate Hamiltonian
perturbation, $\,H_d\rightarrow H_d+x_n(t)D_n\,$, of ``D'' by ``B''.
What is for $\,y_n(t)\,$, they enter (\ref{se}) and (\ref{se1}) like
test functions conjugated with variables $\,D_n\,$. Therefore one can
say that $\,y_n(t)\,$ describe observation of ``D'' by ``B''. But any
thing under observation affects the observer. Hence, in other words,
$\,y_n(t)\,$ represent an opposite action of ``D'' onto ``B''.
Importantly, this passes without self-action of ``D'', which is the
reason for peculiarity of random processes $\,y_n(t)\,$: they are
null by themselves ($\,\left\langle\left\langle
y_{n_1}(t_1)...\,y_{n_k}(t_k)\right\rangle\right\rangle =0\,$)
although possess non-zero cross-correlations with $\,x_n(t)\,$
\cite{i1,i2,i3,i4,i6,i5}. Such correlations are responsible for
energy dissipation in ``D'' and similar statistical effects.

Quantitatively, full statistics of $\,x_n(t)\,$ and $\,y_n(t)\,$ is
determined by separate evolution of ``B'' under perturbations of its
Hamiltonian, $\,H_b\,\rightarrow\,H_b+f_n(t)B_n\,$, by arbitrary
time-varying forces $\,f_n(t)\,$ \cite{i1,i2,i3,i4,i6}. In this
section, let $\,\Gamma \equiv \Gamma_b\,$, and
$\,B_n(t^{\prime},f,\Gamma)\,$ be values of the phase functions
$\,B_n\,$ considered at time $\,t^{\prime}\,$ as functionals of the
forces and functions of current ``B'''s state $\,\Gamma=\Gamma(t)\,$
at time $\,t\,$. Then characteristic functional of $\,x_n(t)\,$ and
$\,y_n(t)\,$ is
\[
\begin{array}{c}
\Xi \{u(\tau),\,f(\tau)\}\equiv
\end{array}
\]
\[
\begin{array}{c}
\equiv\, \left\langle\left\langle \exp\int_{t>t^{\prime}}
[\,u_n(t^{\prime})x_n(t^{\prime})+
f_n(t^{\prime})y_n(t^{\prime})]\,dt^{\prime}\right\rangle
\right\rangle=\,\,
\end{array}
\]
\begin{equation}
\begin{array}{c}
=\text{Tr\,}_b\,\rho _b(t,f,\Gamma)\exp\left\{\int_{t>t^{\prime}}
u_n(t^{\prime})B_n(t^{\prime},f,\Gamma)dt^{\prime}\right\}
\label{cf1}
\end{array}
\end{equation}
with $\,\rho _b(t,f,\Gamma)\,$ being current ``B'''s distribution
function. Since $\,B_n(t^{\prime}=t,f,\Gamma)\equiv B_n(\Gamma)\,$,
the expression under the trace in (\ref{cf1}),
\[
\begin{array}{c}
\widetilde{\rho}_b\,\equiv \,\rho
_b(t,f,\Gamma)\,\exp\left\{\int_{t>t^{\prime}}
u_n(t^{\prime})B_n(t^{\prime},f,\Gamma)\,dt^{\prime}\right\}\,\,\,,
\label{cdf}
\end{array}
\]
satisfies the differential equation
\begin{equation}
d\widetilde{\rho}_b/dt=\left[u_n(t)B_n\,
+\mathcal{L}\left(H_b+f_n(t)B_n\right)\right
]\widetilde{\rho}_b\,\,\,,\label{te}
\end{equation}
quite similar to (\ref{se1}), and CF (\ref{cf1}) can be evaluated by
solving this equation:
\begin{equation}
\begin{array}{c}
\Xi\{u(\tau),\,f(\tau)\}\,=\,\text{Tr}\,_b\,\,
\widetilde{\rho}_b\,\label{cf2}
\end{array}
\end{equation}
Variational differentiations of (\ref{cf1}) produce the identities
\begin{equation}
\begin{array}{c}
\left\langle\left\langle \,\prod _jx(t_j)\prod _my(\tau
_m)\,\right\rangle\right\rangle = \label{cors0}
\end{array}
\end{equation}
\[
=\left [ \prod _m\frac {\delta }{\delta f(\tau _m)}\,\,
\text{Tr}\,_b\,\,\rho _b(t,f,\Gamma)\prod_j B(t_j,f,\Gamma)\, \right
] _{f=0}
\]
clearly explaining the peculiarity of $\,y_n(t)\,$. Besides,
(\ref{cors0}) shows the nullity of any cross-correlations between
$\,y_n(\tau)\,$ and earlier $\,x_n(t^{\prime}\leq \tau)\,$
\cite{i1,i2,i3,i4,i6,i5}, which is consequence of the causality
principle (none perturbation of ``D'' by ``B'' can depend on future
perturbations of ``B'' by ``D'').

{\bf\,IV. Fluctuation-dissipation relations.\,} The phase volume
conservation and generic time-reversal and time-translation
symmetries of Hamiltonian mechanics result in the Onsager reciprocity
relations, Kubo formulas, fluctuation-dissipation theorems \cite{isi}
and other ``fluctuation-dissipation relations'' (FDR)
\cite{bk3,bk2,bk1,bk5}.

In \cite{i6,i5} general quantum FDR were reconsidered in terms of the
stochastic representation. To exploit their classical limit, let us
assume, without loss of generality, that\, (i)
$\,f_n(-\infty)=f_n(+\infty)=0\,$,\,\, (ii) $\,B_n\,$ are chosen so
that their unperturbed mean values are zeroes (i.e.
$\,\left\langle\left\langle x_n(t)\right\rangle\right\rangle =0\,$),
and\, (iii) $\,B_n\,$ possess definite time-reversal parities:
$\,B_n(q,-p)=\epsilon _n B_n(q,p)\,$ with $\,\epsilon _n=\pm 1\,$.
Besides, assume, with a loss of generality, that\, (iv) the past
initial distribution function of ``B'' (before switching-on the
``D''-``B'' interaction) was the canonical one,
$\,\propto\exp(-H_b/T)\,$, and\, (v) $\,H_b\,$ is even: $\,H_b(q,-p)=
H_b(q,p)\,$. Then the classical generating FDR \cite{bk3,bk2,bk1}
yield
\begin{equation}
\Xi\left\{u(\tau)-\frac 1T \frac {df(\tau)}{d\tau
},\,f(\tau)\right\}\,= \Xi\left\{\epsilon u(-\tau),\,\epsilon
f(-\tau)\right\}\,\label{fdr}
\end{equation}
The same can be expressed \cite{i6} by the equalities
\begin{equation}
\begin{array}{c}
\epsilon_n x_n(-\tau)\,\asymp \, x_n(\tau)\,\,\,,\\
\epsilon_n y_n(-\tau)\,\asymp \,
y_n(\tau)+T^{-1}\,dx_n(\tau)/d\tau\,\,, \label{xyfdr}
\end{array}
\end{equation}
where symbol $\,\asymp \,$ means statistical equivalence.

For example, averaging the product of two lines of (\ref{xyfdr})
taken with different arguments, it is easy to obtain such
second-order relation:
\begin{equation}
K_{jm}^{xy}(\tau )=\frac {\theta(\tau)}{T} \frac {d}{d\tau}\,
\,K_{jm}^{xx}(\tau )\,\,,\label{fdt}
\end{equation}
where $\,\theta(\tau)\,$ is the Heavyside step function,
\[
\begin{array}{c}
K_{jm}^{xx}(\tau )\equiv \langle\langle x_j(\tau
)x_m(0)\rangle\rangle \,\,,\,\, K_{jm}^{xy}(\tau )\equiv
\langle\langle x_j(\tau )y_m(0)\rangle\rangle\,\,,
\end{array}
\]
and the causality principle is accounted for as prescribed by
(\ref{cors0}).

{\bf\,V. Distribution function.\,} Come back to ``D'' as described by
the Eqs.\ref{sr},\ref{se},\ref{qcf} and \ref{se1}, using
$\,\Gamma\equiv \Gamma_d\equiv \{q,p\}\,$ as notation for complete
set of ``D'''s variables.

Equation (\ref{se}) can be viewed as generating equation for CF of
variables $\,D_n\,$ in the system with Hamiltonian
$\,H_d+x_n(t)D_n\,$. At that, as we already mentioned, $\,y_n(t)\,$
play the role of test functions conjugated with $\,D_n\,$, while
$\,x_n(t)\,$ are external forces. This picture is described by
Hamilton equations and Liouville equation as follow:
\begin{eqnarray}
d\Gamma(t)/dt\,=\,-[\mathcal{L}\left(H_d+x_n(t)
D_n\right)\Gamma](t)\,\,,\label{he}\\
d\breve{\rho}/dt\,=\,\mathcal{L}\left(H_d+x_n(t)D_n\right)
\breve{\rho}\,\,\,\,\,\,\,\,\,\,\,\,\,\,\,\label{le0}
\end{eqnarray}

Below, let $\Gamma(t)=\Gamma(t,x,\Gamma,\theta)\,$ denote solution of
Eq.\ref{he} with initial condition $\,\Gamma(t=\theta)=\Gamma \,$.
Besides, define $\,\breve{\rho}(t,x,\Gamma)\,$ be solution of
Eq.\ref{le0} under condition
$\,\breve{\rho}(t_0,x,\Gamma)=\rho_{d0}(\Gamma)\,$, where
$\,\rho_{d0}\,$ is $\,\Gamma $'s distribution at past initial time
moment, $\,t_0\,$. Formally, $\,t_0\,$ is the time when the
``D''-``B'' interaction was switched-on. Direct solution of
(\ref{le0}) reads
\begin{equation}
\begin{array}{c}
\breve{\rho}(t,x,\Gamma)=\int\delta\{\Gamma
-\Gamma(t,x,\Gamma_0,t_0)\}\,\rho_{d0}(\Gamma_0)d\Gamma_0
\,=\\=\int\delta\{\Gamma_0
-\Gamma(t_0,x,\Gamma,t)\}\,\rho_{d0}(\Gamma_0)d\Gamma_0=\\
=\rho_{d0}(\Gamma(t_0,x,\Gamma,t)) \,\,\,, \label{df0}
\end{array}
\end{equation}
where $\,\delta\{...\}\,$ means delta-function in the phase space and
$\,\Gamma_0\,$ the initial state. At that, the group property of
$\,\Gamma$'s transformations from one time point to another:
\begin{equation}
\begin{array}{c}
\Gamma(t^{\prime},x,\Gamma(t,x,\Gamma_0,t_0),t)=
\Gamma(t^{\prime},x,\Gamma_0,t_0)\,\,\,,\label{gp}
\end{array}
\end{equation}
and the Liouville theorem about phase volume conservation were taken
into account.

In these designations, solution of Eq.\ref{se} looks as
\begin{equation}
\begin{array}{c}
\widetilde{\rho}_d\,=\,\breve{\rho}(t,x,\Gamma)
\,\exp\left\{\int_{t_0}^t y_n(t^{\prime})D_n(t^{\prime},x,\Gamma,t)\,
dt^{\prime}\right\}\,\,\label{df}
\end{array}
\end{equation}
with $\,D_n(t^{\prime},x,\Gamma,t)\equiv
D_n(\Gamma(t^{\prime},x,\Gamma,t))\,$.

Since $\,y_n(t)\,$ are null by themselves and null in conjunction
with any earlier $\,x_n(t^{\prime}\leq t)\,$, while
$\,\breve{\rho}(t,x,\Gamma)\,$ depends on $\,x_n(t^{\prime}< t)\,$
only, and $\,\Gamma(t^{\prime},x,\Gamma,t)\,$ depend on $\,x_n(\min
(t,t^{\prime})< t^{\prime\prime}< \max(t,t^{\prime}))\,$ only, one
can replace the upper integration limit in (\ref{df}) by any value
$\,>t\,$, in particular, by $\,\infty\,$. Then the exponent in
(\ref{df}) transforms into the statistically equivalent functional
\begin{equation}
S_t\{x,y,\Gamma\}\,\equiv\,\exp\left[\,\int
y_n(t^{\prime})D_n(\Gamma(t^{\prime},x,\Gamma,t))
\,dt^{\prime}\,\right]\,
\label{S}
\end{equation}
After this replacement, substitution of (\ref{df}) to (\ref{sr}),
with use of identities (\ref{df0}) and (\ref{gp}), yields
\begin{equation}
\begin{array}{c}
\rho_d(t,\Gamma)\,=\,\langle\langle\,\breve{\rho}(t,x,\Gamma)
\,S_t\{x,y,\Gamma\}\,\rangle\rangle\,\,=\label{rod}
\end{array}
\end{equation}
\[
\begin{array}{c}
=\int \langle\langle\,\delta\{\Gamma
-\Gamma(t,x,\Gamma_0,t_0)\}\,S_{t_0}\{x,y,\Gamma_0\}\,\rangle\rangle
\,\rho_{d0}(\Gamma_0)d\Gamma_0
\end{array}
\]

Alternatively, by averaging directly formal operator solution of
Eq.\ref{se}, one obtains
\begin{equation}
\begin{array}{c}
\rho_d\,=\,\widehat{\Theta}\,\exp\left\{\int_{t_0}^t
\mathcal{L}(H_d)\,dt^{\prime}\right\}\,
\Xi\{\mathcal{L}(D),D\}\,\rho_{d0}\,\,\,,\label{ddf}
\end{array}
\end{equation}
where $\,\widehat{\Theta}\,$ symbolizes chronological ordering of the
following operator expression (that is ordering with respect to
imaginary time argument of $\,H_d\,$ and $\,D_n\,$).

{\bf\,VI. Fluctuation statistics.\,} Similarly to preceding section,
consider Eq.\ref{se1} as generating equation for joint CF of
variables $\,Q_j\,$ and $\,D_n\,$.

Now, express solutions of (\ref{he}) and (\ref{le0}) through ``D'''s
state at arbitrary fixed time moment $\,\theta\,$ which is different
from $\,t\,$, that is solve (\ref{he}) and (\ref{le0}) under initial
condition $\,\Gamma(t^{\prime}=\theta)=\Gamma\,$ (thus
$\,\Gamma(t^{\prime}=\theta,x,\Gamma,\theta)=\Gamma\,$). Then
solution of Eq.\ref{se1} can be implicitly formulated as
\begin{equation}
\begin{array}{c}
\widetilde{\rho}_d(t,\Gamma(t,x,\Gamma,\theta))=
\,\breve{\rho}(\theta,x,\Gamma)\times\,\\\exp
\left\{\int_{t>t^{\prime}}
[v_j(t^{\prime})Q_j(t^{\prime},x,\Gamma,\theta)
+y_n(t^{\prime})D_n(t^{\prime},x,\Gamma,\theta)]\,dt^{\prime}\right\}
\label{ddf1}
\end{array}
\end{equation}
with $\,Q_j(t^{\prime},x,\Gamma,\theta)\equiv $
$Q_j(\Gamma(t^{\prime},x,\Gamma,\theta))\,$. Substituting
(\ref{ddf1}) to (\ref{qcf}) and taking into account the Liouville
theorem (the phase volume conservation under arbitrary Hamiltonian
evolution), at $\,t\rightarrow\infty\,$ we have
\begin{widetext}
\begin{equation}
\begin{array}{c}
\left\langle \exp\left\{\int
\,v_j(t^{\prime})Q_j(t^{\prime})\,dt^{\prime}\right\} \right\rangle =
\left\langle\left\langle\text{Tr\,}_d\,\exp\left\{\int
v_j(t^{\prime})Q_j(t^{\prime},x,\Gamma,\theta)\,
dt^{\prime}\right\}\breve{\rho}(\theta,x,\Gamma)\,S_{\theta}\{x,y,\Gamma\}
\right\rangle\right\rangle\, \label{fcf}
\end{array}
\end{equation}

In terms of various statistical moments of variables $\,Q_j\,$
(omitting their indices)
\begin{equation}
\left\langle\, Q(t_1)...\,Q(t_k)\,\right\rangle
=\text{Tr\,}_d\left\langle\left\langle
\,Q(t_1,x,\Gamma,\theta)...\,Q(t_k,x,\Gamma,\theta)\,
\breve{\rho}(\theta,x,\Gamma)\,S_{\theta}\{x,y,\Gamma\}
\,\right\rangle\right\rangle   \label{moms}
\end{equation}
In particular, if $\,\theta\rightarrow t_0\,$ then
$\,\breve{\rho}(\theta,x,\Gamma)\,$ turns into the initial
distribution, $\,\rho_{d0}(\Gamma)\,$, definitively independent on
$\,x_n(t)\,$:
\begin{equation}
\left\langle\, Q(t_1)...\,Q(t_k)\,\right\rangle
=\text{Tr\,}_d\,\,\rho_{d0}(\Gamma)\left\langle\left\langle
\,Q(t_1,x,\Gamma,t_0)...\,Q(t_k,x,\Gamma,t_0)\,S_{t_0}\{x,y,\Gamma\}
\,\right\rangle\right\rangle \label{moms0}
\end{equation}
with $\,\Gamma\,$ representing the initial state $\,\Gamma_0\,$. The
same expression results from (\ref{moms}) after substitution of
(\ref{df0}) and (\ref{gp}).

Alternatively, quite similarly to (\ref{ddf}),
\begin{equation}
\begin{array}{c}
\left\langle \exp\left\{\int
\,v_j(t^{\prime})Q_j(t^{\prime})\,dt^{\prime}\right\} \right\rangle
=\text{Tr\,}_d\,\widehat{\Theta}\,\exp\{\,\int
[v_j(t^{\prime})Q_j+\mathcal{L}(H_d)]\,dt^{\prime}\}\,
\,\Xi\{\mathcal{L}(D),D\}\,\rho_{d0}
\end{array}
\end{equation} \label{cdf}
The functional $\,\Xi\,$ here, defined by (\ref{cf1}), at once
accumulates all information about ``B'' which must be used when
evaluating (\ref{fcf})-(\ref{moms0}).
\end{widetext}

{\bf\,VII. Self-interaction through environment and ``scattering
operator''.\,} It is useful to emphasize rather interesting
resemblance between Eq.\ref{moms} or Eq.\ref{moms0} and expressions
for scattering amplitudes, Green functions, etc., in quantum theory
of fields and many-particle systems (see e.g. \cite{blp,lp}). If draw
an analogy from $\,Q_j\,$ and $\,x_n(t)\,$ to electron operators and
radiation field, respectively, then the averages
$\,\left\langle\left\langle
Q(t_1,x,\Gamma,t_0)...Q(t_k,x,\Gamma,t_0)\,\right\rangle\right\rangle\,$
correspond to lowest-order perturbation approximation, while
$\,\left\langle\left\langle
Q(t_1,x,\Gamma,t_0)...Q(t_k,x,\Gamma,t_0)\,S_{t_0}\{x,y,\Gamma\}
\,\right\rangle\right\rangle\,$ in Eq.\ref{moms0} exactly summarizes
all the orders of ``D'''s interaction with its environment. The
analogy continues in that the ``complete multiple scattering
operator'' $\,S_{t_0}\{x,y,\Gamma\}\,$ by itself behaves like unity:
\begin{equation}
\begin{array}{c}
\left\langle\left\langle \,S_{t_0}\{x,y,\Gamma\}
\,\right\rangle\right\rangle \,=1\,
\end{array} \label{unit}
\end{equation}
This identity clearly follows from Eq.\ref{rod} at $\,t\rightarrow
t_0\,$ and is easy explainable if notice that in any term of
$\,\,S_{t_0}\{x,y,\Gamma\}$'s series expansion over
$\,y_n(t^{\prime})\,$ and $\,x_n(t^{\prime})\,$ most late time
argument belongs to some of $\,y$'s.

According to (\ref{cf1}) and (\ref{cors0}), separately $\,x_n(t)\,$
are nothing but noise of free unperturbed environment, like ``zero,
or vacuum, fluctuations''. However, along with $\,y_n(t)\,$ in
$\,S_{t_0}\{x,y,\Gamma\}\,$ they represent actual noise of the
environment, including its directional response to the system's
motion, in the form of both renormalization of primordial ``D'''s
dynamical properties and appearance of new ones: relaxation,
``spectral lines broadening'', etc.

{\bf\,VIII. State-dependent noise and the fiction of friction.\,} In
\cite{i1,i2,i3} the words ``Langevin equation'' were addressed to
objects like (\ref{se}) or (\ref{se1}) which emerged as stochastic
extensions of the Liouville equation for probability measure of
$\,\Gamma\,$. In usual sense, Langevin equations must be a stochastic
extension of the Hamilton equations for $\,\Gamma\,$ themselves.
Besides, one would want these equations to involve some ``realistic''
noises only but not auxiliary ``ghost'' noises like $\,y_n(t)\,$. The
latter requirement means that desirable equations are certainly not
literal consequence of the basic Eqs.\ref{se} and \ref{se1}. Instead,
Langevin equations must be especially constructed as their exact
statistical equivalent (or at least close approximate one).

In should be underlined that, at such target setting, a ``size'' of
system ``B'' is insignificant (no matter e.g. is a Brownian particle
macroscopic or as small as molecules).

With the formulated purpose, let us return to Eq.\ref{moms0},
choosing arbitrary functions $\,Q_j(\Gamma)\,$ as delta-functions
$\,\delta\{\Gamma-\gamma\}\,$ and their index as time. Then
Eq.\ref{moms0} produces
\begin{widetext}
\begin{equation}
\begin{array}{c}
W\{\gamma\}\equiv \left\langle\,\prod_t\delta\{\Gamma(t)-\gamma(t)\}
\,\right\rangle =\int\,\left\langle\left\langle
\,\prod_t\delta\{\Gamma(t,x,\Gamma_0,t_0)-\gamma(t)\}\,\exp\left\{\int
y(t)D(\gamma(t))\,dt\right\}\,\right\rangle\right\rangle\,
\rho_{d0}(\Gamma_0)d\Gamma_0 \label{pf}\\\,
\end{array}
\end{equation}
which represents probability density functional for the whole
system's trajectory. Here all non-principal indices are omitted, and
the delta-functions have allowed to replace
$\,D(\Gamma(t,x,\Gamma_0,t_0))\,$ in the exponent by
$\,D(\gamma(t))\,$.

\,\,

\,\,

The simplest construction of Langevin equations follows directly from
careful ``visual'' investigation of Eq.\ref{pf}. This shows that
Eq.\ref{pf} can be rewritten as
\begin{equation}
\begin{array}{c}
W\{\gamma\}\,=\,\int\,\left\langle\left\langle
\,\prod_t\delta\{\Gamma(t,z,\Gamma_0,t_0)-\gamma(t)\}
\,\right\rangle\right\rangle^{\gamma}\,
\rho_{d0}(\Gamma_0)d\Gamma_0\,\,\,, \label{pf1}
\end{array}
\end{equation}
with new random forces $\,z_n(t)\,$ in place of $\,x_n(t)\,$, if
conditional statistics of $\,z_n(t)\,$ is defined by formulas
\begin{equation}
\begin{array}{c}
\left\langle\left\langle\, z(t_1)...\,z(t_k)\,
\right\rangle\right\rangle^{\gamma} \,
\equiv\,\left\langle\left\langle \,x(t_1)...\,x(t_k)\,\exp\left\{\int
y(t)D(\gamma(t))\,dt\right\}\, \,\right\rangle\right\rangle
\label{momz0}
\end{array}
\end{equation}
The brackets $\,\left\langle\left\langle\,
...\,\right\rangle\right\rangle^{\gamma}\,$  here have the sense of
conditional averaging under given system's trajectory $\,\gamma(t)
\,$.

\,\,

\,\,

At that, the role of Langevin equations governing the variables
$\,\Gamma(t)\equiv \Gamma(t,z,\Gamma,t_0)\,$ and $\,Q(t)\equiv
Q(\Gamma(t,z,\Gamma,t_0))\,$ belongs to nothing but merely the
Hamilton equations:
\begin{eqnarray}
d\Gamma(t)/dt\,=\,-[\mathcal{L}\left(H_d+z_n(t)
D_n\right)\Gamma](t)\,\label{hez}
\end{eqnarray}

\,\,
\end{widetext}

Notice that in view of identity (\ref{unit}) the averaging procedure
defined in (\ref{momz0}) automatically satisfies the normalization
condition $\,\langle\langle\,1 \,\rangle\rangle^{\gamma}=1\,$.
Besides, due to the above mentioned statistical peculiarities of
$\,y(t)$'s, result of the averaging always agrees with the causality
principle: the moments (\ref{momz0}) in fact can depend on
$\,\gamma(t)\,$ with $\,t<\max(t_1,...,t_k)\,$ only.

Formally, the two above expressions, (\ref{momz0}) and (\ref{hez}),
already define what can be named ``exactly equivalent Langevinian
form of the stochastic representation''. It clearly emphasizes
statistical nature of dissipation and friction: even if being present
they still are hidden inside $\,z(t)$'s statistics, as
cross-correlations of $\,x(t)$'s with $\,y(t)$'s. To see them
evidently in (\ref{hez}), we have to withdraw them from (\ref{momz0})
in some reasonable approximation, with corresponding redefinition of
noises $\,z_n(t)\,$.

{\bf\,IX. Unbiased noise and Langevin equations.\,} With the above
pointed purpose, first, assume, naturally and without loss of
generality, that $\,\langle x\rangle =0\,$. Then desired dissipative
contributions to (\ref{hez}), together with the renormalization
corrections of non-dissipative terms, can be identified among mean
values of $\,z(t)$'s.

Second, consider cumulants (semi-invariants) $\,\,
\kappa_{\alpha\beta}\,\equiv\,\langle\langle\,
x_1,...,x_{\alpha},\,y_1,...,y_{\beta}\,\rangle\rangle \,$. For
brevity, here the subscripts unify indices and time, and commas do
emphasize that comma-separated multipliers are subject to purely
irreducible correlation of $\,(\alpha +\beta)$-th order (``Malakhov's
cumulant brackets''). Then CF (\ref{cf1}) can be symbolically written
as
\begin{widetext}
\[
\Xi\{u,f\}\,=\,\exp [\,\kappa\{u,f\}\,]\,\equiv\,
\exp\left[\,\sum_{\alpha=2}^{\infty}\kappa_{\alpha 0}\frac
{u^{\alpha}}{\alpha !}\,+\,\sum_{\alpha,\beta=1}^{\infty}
\kappa_{\alpha\beta}\frac
{u^{\alpha}f^{\beta}}{\alpha!\,\beta!}\right]
\]
(according to our assumption, $\,\kappa_{10}=\langle x\rangle =0\,$).
Decompose it into two multipliers:
\begin{equation}
\Xi\{u,f\}=\overline{\Xi}\{u,f\}\,\widetilde{\Xi}\{u,f\}\,\,,\,\,\,
\overline{\Xi}\{u,f\}\equiv\exp\left[\sum_{\beta=1}^{\infty}\kappa_{1
\beta}\frac {u\,f^{\beta}}{\beta!}\right]\,\,,\,\,\,
\widetilde{\Xi}\{u,f\}\equiv\exp\left[\sum_{\alpha=2}^{\infty}
\sum_{\beta=0}^{\infty}\kappa_{\alpha \beta}\frac
{u^{\alpha}f^{\beta}}{\alpha!\,\beta!}\right]  \label{dec}
\end{equation}
Correspondingly to this factorization of CF, both the original
noises, $\,x(t)$'s and $\,y(t)$'s, divides into two components:
$\,\,x=\overline{x}+\widetilde{x}\,$ and
$\,y=\overline{y}+\widetilde{y}\,$, where two pairs
$\,\{\widetilde{x},\widetilde{y}\}\,$ and
$\,\{\overline{x},\overline{y}\}\,$ are mutually statistically
independent.

It is easy to prove that for arbitrary functional
$\,\Phi(\overline{x})\,$ and arbitrary function $\,f\,$ the equality
holds as follows:
\begin{equation}
\langle\langle\,
\Phi(\overline{x})\,\exp(f\overline{y})\,\rangle\rangle =
\Phi\left(\overline{X}(f)\right)\,\,,\,\,\,\overline{X}(f)\equiv
\sum_{\beta=1}^{\infty}\kappa_{1\beta}\,f^{\beta}/\beta!\,\label{mv}
\end{equation}
This is because the pair $\,\{\overline{x},\overline{y}\}\,$, in
accordance with (\ref{cors0}), describes merely conditional mean
value of ``B'''s response to its perturbation by forces $\,f\,$.
Applying the decomposition (\ref{dec}) to (\ref{pf}), with the help
of (\ref{mv}) we obtain

\begin{equation}
\begin{array}{c}
W\{\gamma\}=\,\int \left\langle\left\langle\,\,
\prod_t\delta\{\Gamma(t,\overline{X}(D(\gamma))+\widetilde{x},
\Gamma_0,t_0)-\gamma(t)\}\,\exp\left\{\int
\widetilde{y}(t)D(\gamma(t))\,dt\right\}\,\right\rangle\right\rangle
\,\rho_{d0}(\Gamma_0) \,d\Gamma_0\,\,\label{pf2}
\\\,
\end{array}
\end{equation}
The mean response $\,\overline{X}(f)\,$, defined by (\ref{mv}), with
$\,f=D(\gamma)\,$, after restoration of its temporal index, reads
\begin{equation}
\overline{X}(t,f)=\frac {\delta}{\delta
u(t)}\,\ln\Xi\{u,f\}\,|_{u=0}\,=\int\langle\langle
x(t),y(t_1)\rangle\rangle\, f(t_1)dt_1+\frac 12\int\int\langle\langle
x(t),y(t_1),y(t_2)\rangle\rangle\,f(t_1)f(t_2)dt_1dt_2\,+...\,\,,\label{mx}
\\\,\,
\end{equation}
where because of (\ref{cors0}) all integrals are in fact taken over
$\,t_j<t\,$. It is useful to notice also that, due to the causality,
Jacobian of mutual transformations between $\,\gamma\,$ and
$\,\Gamma\,$ is unit.

Scanning (\ref{pf2}) in comparison with (\ref{pf}) and (\ref{pf1}),
one evidently comes to another form of the probability functional:

\begin{equation}
\begin{array}{c}
W\{\gamma\}\,=\,\int\,\left\langle\left\langle
\,\prod_t\delta\{\,\Gamma(t,\overline{X}(D(\gamma))+\widetilde{z}\,,
\Gamma_0,t_0)-\gamma(t)\} \,\right\rangle\right\rangle^{\gamma}\,
\rho_{d0}(\Gamma_0)d\Gamma_0\,\,\,, \label{pf3}
\\\,\,
\end{array}
\end{equation}
where statistics of renormalized (in fact merely biased) noises
$\,\widetilde{z}(t)\,$ is now described by

\begin{equation}
\begin{array}{c}
\left\langle\left\langle\,
\widetilde{z}(t_1)...\,\widetilde{z}(t_k)\,
\right\rangle\right\rangle^{\gamma} \,
\equiv\,\left\langle\left\langle
\,\widetilde{x}(t_1)...\,\widetilde{x}(t_k)\,\exp\left\{\int
\widetilde{y}(t)D(\gamma(t))\,dt\right\}
\,\right\rangle\right\rangle\,\,\label{momz1}
\\\,
\end{array}
\end{equation}
At that, correspondingly to (\ref{pf3}),
$\,\Gamma(t)=\Gamma(t,\overline{X}(D(\Gamma))+\widetilde{z},\Gamma_0,t_0)\,$,
that is stochastic Hamilton equations (\ref{hez}) change to the
stochastic integro-differential equations
\begin{eqnarray}
d\Gamma(t)/dt\,=\,-\,[\,\mathcal{L}(H_d)\Gamma\,](t)-
(\overline{X}_n(t,D(\Gamma))+\widetilde{z}_n(t))\,
[\mathcal{L}(D_n)\,\Gamma](t)\,\,\label{hez1}
\end{eqnarray}
\end{widetext}
As prescribed by (\ref{dec}) and (\ref{momz1}), here the noises
$\,\widetilde{z}_n(t)\,$ have certainly zero mean values, while any
dissipative effects of interaction with ``B'' are separated in
$\,\overline{X}_n(t,D(\Gamma))\,$. Hence, Eqs.\ref{hez1} can be by
now enough surely named ``Langevin equations''.

{\bf\,X. Discussion.\,} Of course, the above result is rather trivial
one. However, from the point of view of applications and practical
computability, it is not quite satisfactory. The matter is that
numeric modeling of noise essentially conditioned by the system it
drives is generally difficult task. It would be better if the noise
was reduced to unconditioned random quantities, for example,
\begin{equation}
\begin{array}{c}
\widetilde{z}(t)=z^{(0)}(t)+
\int\,z^{(1)}(t,t_1)\,f(t_1)\,dt_1\,+\\+\,\frac
12\int\int\,z^{(2)}(t,t_1,t_2)\,f(t_1)f(t_2)\,dt_1dt_2\,+...
\,\,\,,\label{expan}
\end{array}
\end{equation}
where $\,z^{(0)}(t)=\widetilde{x}(t)=x(t)\,$ is unperturbed noise,
$\,z^{(1)}(t,t_1)\,$ represents stochastic linear response of ``B''
to its perturbation, etc., and all $\,z^{(n)}\,$ are some
zero-average random functions independent on the forces. In
particular, $\,z^{(1)}(t,t_1)\,$ includes fluctuations in linear
friction (whose average was contained in first term of (\ref{mx})).

The only situation when Eqs.\ref{hez1} finalize the analysis is when
the noises $\,\widetilde{z}(t)\,$ are state-independent, that is
$\,z^{(n)}=0\,$ for all $\,n>0\,$. But this is unlikely realistic
situation since in general it is forbidden by restrictions which
follow from the phase volume conservation and microscopic
reversibility. For concreteness, if ``B'' is equilibrium thermal bath
(thermostat), these restrictions are expressed by FDR (\ref{fdr})
\cite{bk3,bk2,bk1,bk5} or equivalently (\ref{xyfdr}) (notice that FDR
for internally non-equilibrium baths also were considered in
\cite{bk3,bk2}). If noises $\,\widetilde{z}(t)\,$ are indeed
state-independent, this means that $\,\kappa_{\alpha\beta}=0\,$ for
all $\alpha\geq 2\,$ and $\,\beta\geq 1\,$. Then the second row from
(\ref{xyfdr}) clearly implies that in such case the equalities
$\,\kappa_{\alpha 0}=0\,$ also should hold for all $\alpha\geq 3\,$.
In other words, the noise $\,\widetilde{z}(t)\,$ can be purely
state-independent only when it is purely Gaussian. Moreover, then the
same FDR prescribe that $\,\kappa_{1\beta}=0\,$ for all $\,\beta\geq
2\,$, that is average response of ``B'' is purely linear.

Thus we come to the trite ``linear Gaussian thermostat'' when
Eqs.\ref{momz1} and \ref{mx} reduce to
\begin{equation}
\begin{array}{c}
\overline{X}_n(t,D(\Gamma))\,=\,-\,
K_{nm}^{xx}(0)D_m(\Gamma(t))/T\,\,+\,\,\label{lin}
\end{array}
\end{equation}
\[
+\,\frac 1T\int^{t}_{-\infty} K_{nm}^{xx}(t-t^{\prime})\,\frac
{d}{dt^{\prime}} \,D_m(\Gamma(t^{\prime}))\,dt^{\prime}\,\,\,,
\]
\[
\begin{array}{c}
\langle\langle\,\, \widetilde{z}_n(t_1)\,\widetilde{z}_m(t_2)\,\,
\rangle\rangle^{\gamma}\,=\,K_{nm}^{xx}(t_1-t_2)\,\\\,
\end{array}
\]
Here FDR (\ref{fdt}) is used, and it is taken in mind that all
higher-order cumulants of $\,\widetilde{z}(t)\,$ are zeros.
Discussion of more interest models will be done elsewhere.

{\bf\,XI. Example: oscillator.\,} Consider nonlinear oscillator,
assuming that ``B'' is ``linear Gaussian thermostat'' while
interaction with it realizes in potential way through two
statistically independent channels as follow:
\[
\begin{array}{c}
H_d=p^2/2m+U_0(q)\,\,,\,\,\,D_1(\Gamma)=-q\,\,,\\
\,\,\,D_2(\Gamma)=q^2/2\,\,,\,\,\,K^{xx}_{12}=K^{xx}_{21}=0
\end{array}
\]

The first channel corresponds to usual thermal excitation, and the
second to thermal parametric fluctuations in frequency of
oscillations. The Eqs.\ref{hez1} and \ref{lin} yield
\begin{equation}
\begin{array}{c}
dq(t)/dt\,=\,p(t)/m\,\,\,,\\\,\label{osc} \\
dp(t)/dt\,=\,-\,dU(q(t))/dq(t)\,+\widetilde{x}_1(t)\,+
\widetilde{x}_2(t)\,q(t)\,-\\-\,\int^t_{-\infty}
K_{11}^{xx}(t-t^{\prime})\,v(t^{\prime})\,dt^{\prime}/T\,-\\
-\,q(t)\int^{t}_{-\infty}
K_{22}^{xx}(t-t^{\prime})\,q(t^{\prime})v(t^{\prime})
\,dt^{\prime}/T\,\,,
\end{array}
\end{equation}
where $\,v(t)\equiv dq(t)/dt\,$ is velocity, $\,\widetilde{x}_n(t)\,$
are mutually independent normal random processes, $\,K_{nn}^{xx}\,$
are their correlators, and
\[
\begin{array}{c}
U(q)\,\equiv\,
U_0(q)\,-K_{11}^{xx}(0)\,q^2/2T\,-K_{22}^{xx}(0)\,q^4/8T
\end{array}
\]
is renormalized potential. Hence, correspondingly, there are two
channels of friction and dissipation, and the friction channel
conjugated with thermal parametric fluctuations is essentially
nonlinear. Similar examples concerning thermal fluctuations in
capacities of electric circuits were considered in \cite{bk4}.

{\bf\,XII. Conclusion.\,} For particular variant of the ``stochastic
representation of deterministic interactions'' concerning classical
Hamiltonian mechanics, we have demonstrated that by request it can be
completely reformulated in terms of ``Langevin equations'' for
internal variables of an open system. These equations are wholly
housed in its own phase space and are free of the peculiar auxiliary
noises $\,y_n(t)\,$, distinctive for initial ``stochastic
representation''. At the same time, $\,y_n(t)\,$ remain useful
undercover instrument, being responsible for conditional statistical
dependence of actual noise on trajectory of the system driven by it.

This Langevinian form of the theory seems more vivid, although,
probably, it will occur less appropriate for practical analysis of
complicated noise statistics. Besides, the original ``Liouvillian
form'' at once covers quantum mechanics as well.

What is for its quantum Langevinian equivalent, still it remains
unexplored. Notice that quantum Langevin equations for important
special case of Gaussian linear thermostat were exhaustively
considered in \cite{ef1}. Of course, more general situations also
were under many considerations (see e.g. \cite{ef}).

But recall that the question under our principal and pragmatic
interest is how much non-Gaussian non-linear generalization of
quantum Langevin equations can be developed if do it wholly within
native Hilbert space of an open system under consideration and with
use of commutative ($\,c$-number valued) noise sources only.

\,\,



\end{document}